\documentclass[superscriptaddress,twocolumn,prb,showpacs,floatfix]{revtex4}
\usepackage[dvips]{graphicx}
\usepackage{amsmath,amsfonts,amssymb,bm,ulem}

\usepackage{graphicx}
\usepackage{epsf}

\newcommand{\B}{{\text{B}}}

\usepackage[usenames]{color}

 \newcommand{\bz}{\mathbf{z}}

    \newcommand{\bx}{\mathbf{x}}

\begin{document}

\title{Magnetotransport in low-density $p$-Si/SiGe heterostructures:
  From metal through hopping insulator to Wigner glass}

\author{I. L.~Drichko}
\email{irina.l.drichko@mail.ioffe.ru}
\affiliation{A. F. Ioffe
Physico-Technical Institute of Russian Academy of Sciences, 194021
St. Petersburg, Russia}
\author{A. M.~Dyakonov}
\affiliation{A. F. Ioffe  Physico-Technical Institute of Russian
Academy of Sciences, 194021 St. Petersburg, Russia}
\author{I. Yu.~Smirnov}
\affiliation{A. F. Ioffe  Physico-Technical Institute of Russian
Academy of Sciences, 194021 St. Petersburg, Russia}
\author{A. V.~Suslov}
\affiliation{National High Magnetic Field Laboratory, Tallahassee,
FL 32310, USA}
\author{Y.~M.~Galperin}
\affiliation{Department of Physics and Center for Advanced Materials
  and Nanotechnology, University of Oslo, PO Box
1048 Blindern, 0316 Oslo, Norway} \affiliation{A. F. Ioffe
Physico-Technical Institute of Russian Academy of Sciences, 194021
St. Petersburg, Russia} \affiliation{Argonne National Laboratory,
9700 S. Cass Av., Argonne, IL 60439, USA}
\author{V.~Vinokur}
\affiliation{Argonne National Laboratory, 9700 S. Cass Av.,
Argonne, IL 60439, USA}
\author{M.~Myronov}
\affiliation{Musashi Institute of Technology, 8-15-1 Todoroki,
Setagaya-ku, Tokyo, Japan}
\author{O. A.~Mironov}
\affiliation{Warwick SEMINANO R$\&$D Centre, University of Warwick
Science Park, Coventry CV4 7EZ, UK}
\author{D. R.~Leadley}
\affiliation{Department of Physics, University of Warwick, Coventry
CV4 7AL, UK}
\date{\today}
\begin{abstract}
We study DC and AC transport in low-density $p-$Si/SiGe
heterostructures at low temperatures and in a broad domain of
magnetic fields up to 18 T. Complex AC conductance is determined
from simultaneous measurement of velocity and attenuation of a
surface acoustic wave propagating in close vicinity of the 2D hole
layer. The observed behaviors of DC and AC conductance are
interpreted as an evolution from metallic conductance at $B=0$
through hopping between localized states in intermediate magnetic
fields (close to the plateau of the integer quantum Hall effect
corresponding to the Landau-level filling factor $\nu$=1) to
formation of the Wigner glass in the extreme quantum limit
($B\gtrsim 14$, $T \lesssim 0.8$ K).
\end{abstract}

 \maketitle

\section{Introduction} \label{intr}
Electron transport through two-dimensional (2D) semiconductor structures
attracts great attention during many years. Firstly, two-dimensional
layers serve as building blocks for devices of modern micro- and
nano-electronics. Secondly, two-dimensional structures allow
addressing several fundamental problems. In particular, they provide
possibilities to analyze interplay between the electron-electron
interaction and disorder, which is the most fundamental problem of
modern condensed matter physics. External magnetic field adds a ``new
dimension'' to these studies. A parallel magnetic field acts mostly
upon spin degrees of freedom causing splitting of the electron levels
corresponding to different spins. A perpendicular magnetic field acts
mostly upon orbital degrees of freedom by shrinking electron wave
functions. As a result, it influences the interaction-disorder
interplay in a controlled way.

The most remarkable magneto-transport phenomena existing only in
2D systems are the integer and fractional quantum Hall effects
(IQHE and FQHE, respectively). While the FQHE can be observed only
in very pure samples having extremely high mobility, the IQHE is a
typical feature of relatively disordered structures. Furthermore,
according to conventional picture of the IQHE (see, e.\, g.,
Ref.~\onlinecite{qhe} for a review), it is the disorder
responsible for the ``reservoir" of localized states which
controls the evolution of the chemical potential with variation of
the electron/hole density or magnetic field. The localized states,
in turn, lead to broad plateaus in the dependence of the Hall
component, $\rho_{xy}$,  of the conductance tensor on magnetic
field or electron density, and vanishing of the transverse
component, $\rho_{xx}$, at the plateaus. As usual, we assume that
magnetic field is parallel to the $\bz$-axis while the electric
field is parallel to the $\bx$-axis.

According to quantum mechanics,~\cite{LL} the orbital energy
spectrum of a perfect 2D electron system in a perpendicular magnetic
field consists of discrete Landau levels (LLs), $\varepsilon_n
=\hbar \omega_c(n+1/2)$ where $\omega_c=eB/m^*c$ is the cyclotron
frequency; $B$ is the external magnetic field, $e$ is the electron
charge, $m^*$ is the (cyclotron) effective mass, while $c$ is the
light velocity. The degeneracy factor of the levels is just the
ratio between the sample area, $A$, and the effective  area $2\pi
l_B^2$ occupied by a quantum state. Here $l_B\equiv(\hbar
c/eB)^{1/2}$ is the so-called Landau or magnetic length. The filling
factor, $\nu=2\pi p l_B^2=p c h/ eB$, has a meaning of the ratio of
the electron number to the ``capacity'' of a Landau level. Here $p$
is the sheet electron/hole density. An integer $\nu$ means that an
integer number of LLs are fully occupied and the chemical potential
is located in the gap between them. In the above consideration, we
have defined the filling factor \textit{per electron spin}. The
external field causes Zeeman splitting, $g^*\mu_B B$, of the levels
corresponding to different spins where $\mu_B$ is the Bohr magneton
while $g^*$ is the so-called g-factor. If Zeeman splitting exceeds
the thermal splitting, $k_\B T$, then the spin-split levels are well
resolved. Here $T$ is temperature while $k_\B$ is the Boltzmann
constant.

The domain where $\nu < 1$ and $k_\B T \lesssim \hbar \omega_c$ is
called the \textit{extreme quantum limit} (EQL). In this domain the
electron states are spin-polarized and only the lowest Landau level
is partly occupied.  The magneto-transport in the EQL region is far
from being fully understood. There exist predictions that a 2D
system in such situation behaves as a specific ``Hall insulator''
where the off-diagonal component, $\rho_{xy}$, of the resistivity
tensor keeps its classical values while the diagonal component,
$\rho_{xx}$, diverges at zero temperature.~\cite{insul}

This is in contrast with an ordinary (Anderson or Mott) insulator where
both components diverge. Transverse DC conductance of both the
Hall and an ordinary  insulator at finite temperature is due to
the variable-range
hopping of electrons (holes) between localized states. It turns out
that AC conductivity in this regime is complex, $\sigma^{\text{AC}}
\equiv \sigma_1
 -i\sigma_2$ and $\sigma_2 > \sigma_1 >
\sigma^{\text{DC}}_{xx}=\rho_{xx}/(\rho_{xx}^2+\rho_{xy}^2)$.~\cite{Efros}
This
relation has been experimentally confirmed in GaAs/AlGaAs
heterostructures near the conductivity minima
in the IQHE regime using probeless acoustic
methods to measure AC conductivity.~\cite{Drichko00} The observed large value
of the ratio $\sigma_2 / \sigma_1$ was interpreted as  a
hallmark of hopping conductance since the contribution of extended
carriers to $\sigma_2$ is extremely small in the studied frequency domain.

An alternative scenario for low-temperature behavior of an interacting
2D system is formation of the Wigner
crystal -- periodic distribution of charge carriers.~\cite{Wigner34}
In a pure electron system an interplay
between the kinetic and interaction energy of electrons depends only
on their density. At sufficiently low density, the typical interaction
energy can exceed the Fermi energy of free electrons, and this is just
the domain where the Wigner crystal can be formed.
Electron Wigner crystal was observed for the first
on the surface of liquid He.~\cite{Grimes79} It was also
identified in Si metal-oxide-semiconductor (MOS) structures with low
electron concentration and high mobility, see
Ref.~\onlinecite{Pudalov94} for a review.

External transverse  magnetic field shrinks the electron wave
functions and in this way facilitates formation of the Wigner
crystal. As a result, in the presence of sufficiently strong
magnetic field 2DEG can form the Wigner crystal even at relatively
high electron concentration for which at $B=0$ the electron system
is a liquid.~\cite{Lozovik75} The Wigner crystallization in magnetic
field was studied by many research groups, both experimentally and
theoretically. Most of experiments were done using high-quality
heterostructures such as $n$-GaAs/AlGaAs,~\cite{Jiang92} inversion
high-mobility Si films,~\cite{Dolgopolov92} InGaAs/InP
heterostructures.~\cite{Podor01} The most popular experimental
method here is studies of DC $I-V$
curves.~\cite{Pudalov94,Dolgopolov92,Podor01}

The conventional point of view resulting from the above works is
that in realistic 2D systems the Wigner crystal is strongly
distorted by disorder and consists of correlated regions sometimes
called ``the domains''. The whole structure is pinned by disorder
forming a glass-like system see, e.\,g., Ref.~\onlinecite{Fogler99}.
This \textit{Wigner glass} should exhibit specific nonlinear and
hysteretic response to the applied voltage, which is typical for
pinned interacting random
systems.~\cite{Ioffe87,Feigelman89,Blatter94} In the presence of AC
excitation electrons vibrate around the pinning centers in a
collective fashion forming the so-called \textit{pinning collective
mode} strongly influenced by the magnetic field. This mode was
identified as a specific resonance in the AC conductance observed in
high-mobility $n$ and $p$-GaAs/AlGaAs heterostructures at $\nu
<0.2$, i.\,e., in the EQL. The typical resonant frequencies were of
few GHz.~\cite{Williams91,Ye02,Li00,Yong06}

This work is aimed at studies of DC and AC magnetotransport in
low-density heterostructures $p$-Si/SiGe
($p=8\times 10^{10}$ cm$^{-2}$) both in the IQHE ($\nu =1$) and EQL regimes.
 In this material, the ratio
between the typical hole-hole interaction energy and the Fermi
energy is about 10; an additional advantage is that here formation
of the Wigner crystal is not masked by the liquid phases
corresponding to the fractional quantum Hall effect. In addition to
conventional DC measurements, we measure velocity and attenuation of
a surface acoustic wave (SAW) excited at the surface of a piezoelectric
crystal located close to the 2D hole layer in the heterostructure.
These measurements conducted at different temperatures and magnetic
fields provide a probeless method for studying AC response. This
method allows one determining the \textit{complex} AC conductance,
it has been previously successfully applied to $n$-GaAs/AlGaAs for
identifying of Wigner crystal.~\cite{Paalanen92} In this way we will
demonstrate the evolution from metallic conductance at $B=0$ through
hopping conductance in intermediate magnetic fields to formation of
the Wigner glass in very high magnetic fields.

The paper is organized as follows. In Sec.~\ref{experiment} we report
the procedures of measurement and data handling. The results are
presented  in Sec.~\ref{results} and discussed in
Sec.~\ref{discussion}.

\section{Experiment} \label{experiment}

 \subsection{Procedure} \label{procedure}

\begin{figure}[ht]
\centerline{
\includegraphics[width=\columnwidth]{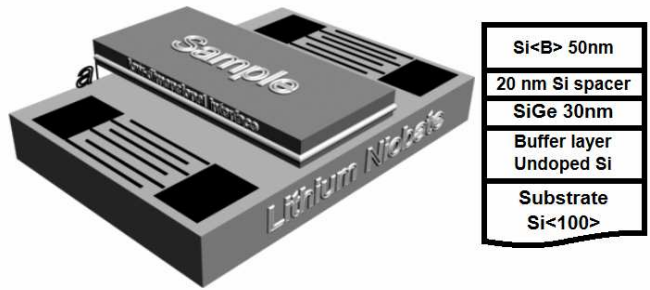}
} \caption{Sketches of the acoustic experiment setup and sample.
\label{Sample}}
\end{figure}

We simultaneously measured attenuation and velocity of SAW in
$p$-Si/SiGe heterostructures with hole density  $p = 8.2\times
10^{10}$~cm$^{-2}$ and mobility $\mu  = 1\times10^{4}$
cm$^2$/V$\cdot$s in external magnetic field up to 18 T and
temperature interval $T=0.3-4.2$ K. The measurements were performed
in the frequency domain $f=18-255$ MHz using the so-called hybrid
method, see, e.\,g., Ref.~\onlinecite{Drichko00}. According to this
method, the SAW was excited by an inter-digital transducer at the
surface of a piezoelectric crystal, LiNbO$_3$, the heterostructure
sample being pressed to the surface as illustrated on
Fig.~\ref{Sample}. The SAW generates a moving profile of electric
field, which penetrates the 2D-interface, causing AC electric
current. The current produces Joule heating, as well as feedback
forces acting upon the elastic medium in the piezoelectric crystal.
These processes result in an additional attenuation, $\Delta\Gamma$,
of the SAW, as well as variation,  $\Delta v$,  of its velocity.
Both effects depend on the conductance of the two-dimensional hole
gas (2DHG) at the 2D interface. Consequently, by simultaneous
measurement of $\Delta\Gamma$ and $\Delta v$ we extract
\textit{complex conductivity}, $\sigma^{\text{AC}} (\omega) \equiv
\sigma_1 (\omega) -i\sigma_2 (\omega)$, of the 2D hole system versus
magnetic field, temperature, and SAW amplitude. This "sandwich"-like
method allows to study non-piezoelectric systems by acoustic
methods.

In addition to acoustic experiments, we measured components
$\rho_{xx}$ and $\rho_{xy}$ of static electrical resistance, as
well as the static current-voltage ($I-V$) curves of a similar
sample in magnetic field up to 18 T and temperatures $T=0.3-2.1$
K.

The polymorphic fully strained heterostructure
Si$\langle$B$\rangle$/Si/SiGe/Si/(001)Si was grown using Solid Source
MBE with e-beam on the
substrate Si (100) (Fig.~\ref{Sample}). It consisted of the 300 nm Si
buffer layer followed by 30 nm
Si$_{0.92}$Ge$_{0.08}$  layer, 20 nm undoped spacer and
50 nm layer of B-doped Si with doping concentration 2.5$\times
10^{18}$~cm$^{-3}$. The 2D interface was located in the
strained Si$_{0.92}$Ge$_{0.08}$ layer, see  Ref.~\onlinecite{Fedina}
for detailed analysis of the sample.

The samples used for acoustic measurements were of the size
0.3$\times$0.5 cm$^2$. The DC measurements were performed using a
satellite sample made from adjacent sector of the same
heterostructure plate and shaped as a Hall bar. Sputtered Al strips
annealed at 500$^\circ$ C for 30 min served as Ohmic potential
contacts, see Ref.~\onlinecite{Agan}.

\subsubsection{Linear regime} \label{linear}

\begin{figure}[ht]
\centerline{
\includegraphics[width=\columnwidth]{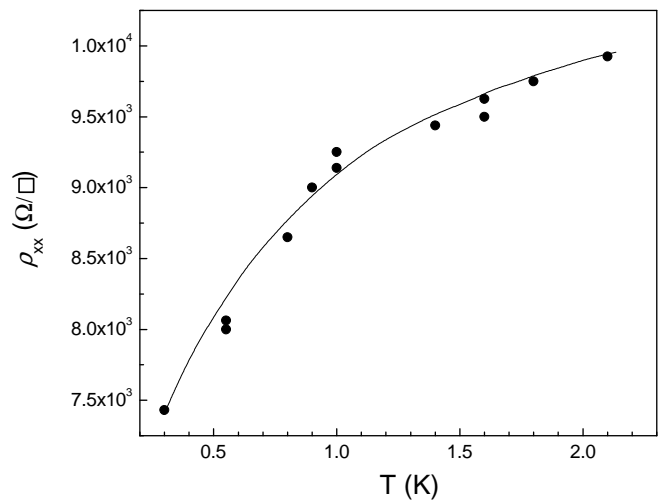}
} \caption{Temperature dependence of static resistivity,
$\rho_{xx}$, $B=0$, $I=10$ nA. The line is a guide to eye. \label{fig:1}}
\end{figure}

\paragraph{DC conductivity.}
Shown in Fig.~\ref{fig:1} is the temperature dependence of the static
resistivity,
$\rho_{xx}$, measured at $B=0$ and small
measurement current (10 nA).
This dependence corresponds to a metal-like behavior, the
resistivity at the lowest temperature (0.3 K) being $(7.5\pm 0.1) \
\text{kOhm}/\Box \approx 0.29 \, h/e^2$. In this regime, the $I-V$
curve remains Ohmic up the current of 300 nA. Thus, one concludes
that at $B=0$ the hole system is in a metallic state even at the
lowest studied temperature.

The components  $\rho_{xx}$ and $\rho_{xy}$ were measured in the
temperature domain  $0.3-2$ K and in magnetic fields up to 18 T
($\rho_{xx}$) and 8 T ($\rho_{xy}$). We were not able to measure
$\rho_{xy}$ in larger magnetic fields because there the  $\rho_{xx}$
turns out to be too large -- 1.4 GOhm at $T=0.3$ K and measurement
current $I=0.5$ nA. At such resistance the
Hall voltage is masked by  potential difference
between the voltage probes along the $\bx$ axis.

\begin{figure}[ht]
\centerline{
\includegraphics[width=\columnwidth]{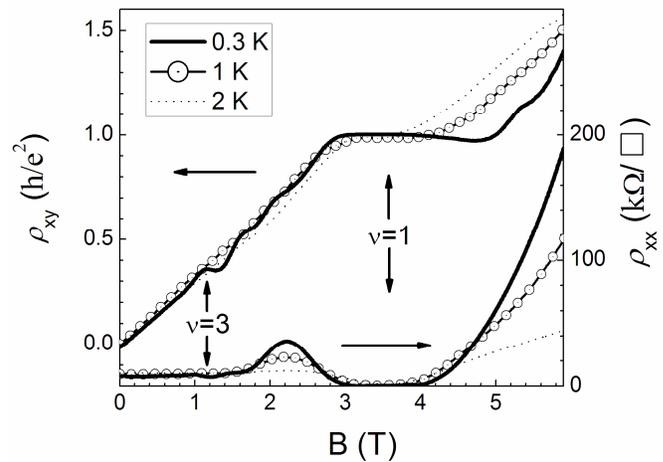}
} \caption{Magnetic field dependences of $\rho_{xx}$ and
$\rho_{xy}$ at
 different temperatures. \label{fig:2}}
\end{figure}

The components of
magnetoresistance versus magnetic field for $B \le 6$ T are shown in
Fig.~\ref{fig:2}.
One can notice  a minimum in $\rho_{xx}(B)$ due to the Shubnikov-de
Haas (SdH) effect at the filling factor $\nu=1$, as well as integer
quantum Hall plateaus in $\rho_{xy}$. There is also a weak deep at
$\nu=3$, which is not clearly seen in Fig.~\ref{fig:2} because of
chosen scale. The absence of the minima for even values of the
filling factor is usual for strained $p$-SiGe, see, e.\,g.,
Ref.~\onlinecite{Coleridge} and references therein. The reason is
the following. While the valence band in $p$-SiGe is 6-fold
degenerate if both spin-orbit interaction and strain in the
quantum-well structure are neglected the spin-orbit interaction plus
strain partly lift the degeneracy leading to the energy separation
of 23~meV  between the heavy and light holes.\cite{Coleridge}
Therefore, the conductivity is maintained by the heavy holes. The
spin splitting of the heavy holes in $p$-SiGe 2D-systems  is
significantly enhanced by exchange interaction. As a result, the
spin splitting turns out to be close to the half of the cyclotron
splitting, $g^*\mu_B B \approx \hbar \omega_c/2$, where
$\omega_c=eB/m^*c$ is the cyclotron frequency. This is why all even
deeps are suppressed. This behavior strongly differs from that
observed in the A$^{\text{III}}$B$^{\text{V}}$ heterostructures.

 At $B \ge 4.5$ T and $T \lesssim $10 K the holes occupy the states only
of lowest spin-splitted band of the 0-Landau level, so the condition
of the quantum limit is fulfilled.

\begin{figure}[ht]
\centerline{
\includegraphics[width=\columnwidth]{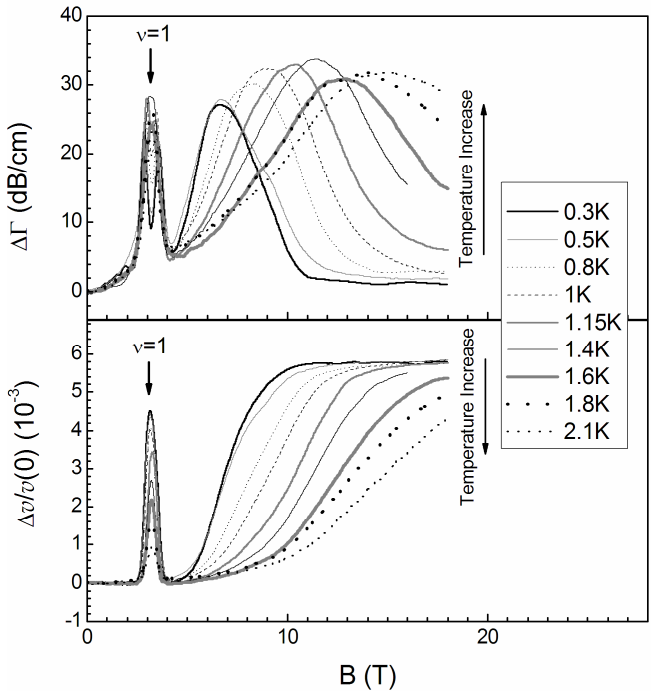}
} \caption{Magnetic field dependences of $\Delta \Gamma$ and
$\Delta
  v/v(0)$ for different temperatures. $f=87.7$ MHz, $p=8.2 \times
  10^{10}$ cm$^{-2}$.
 \label{fig:3}}
\end{figure}

\paragraph{Acoustic properties.}
Shown in Fig.~\ref{fig:3} are magnetic field dependences of the
acoustic attenuation, $\Delta \Gamma (B) \equiv
\Gamma(B)-\Gamma(0)$, and velocity, $\Delta v(B)/v(0)  \equiv [v(B)
- v(0)]/v(0)$ at $f=87.7$ MHz. Here $\Gamma(0)$ and $v(0)$ are the
SAW  attenuation and velocity at $B$=0, respectively.
One can see pronounced extrema in both  $\Delta \Gamma$ and
$\Delta
  v/v(0)$ at the magnetic field corresponding to  $\nu =1$. These extrema
  coincide with the pronounced deep in static $\rho_{xx}$,
  Fig.~\ref{fig:2}. Above  $B \ge 4.5$ T the system is in the
  extreme quantum limit, $\hbar \omega_c \gtrsim T$.
The curves measured at different frequencies (18, 30, 157, and 240
MHz) are similar.~\cite{end1}

\subsubsection{Nonlinear regime} \label {nonlinear}

\begin{figure}[ht]
\centerline{
\includegraphics[width=\columnwidth]{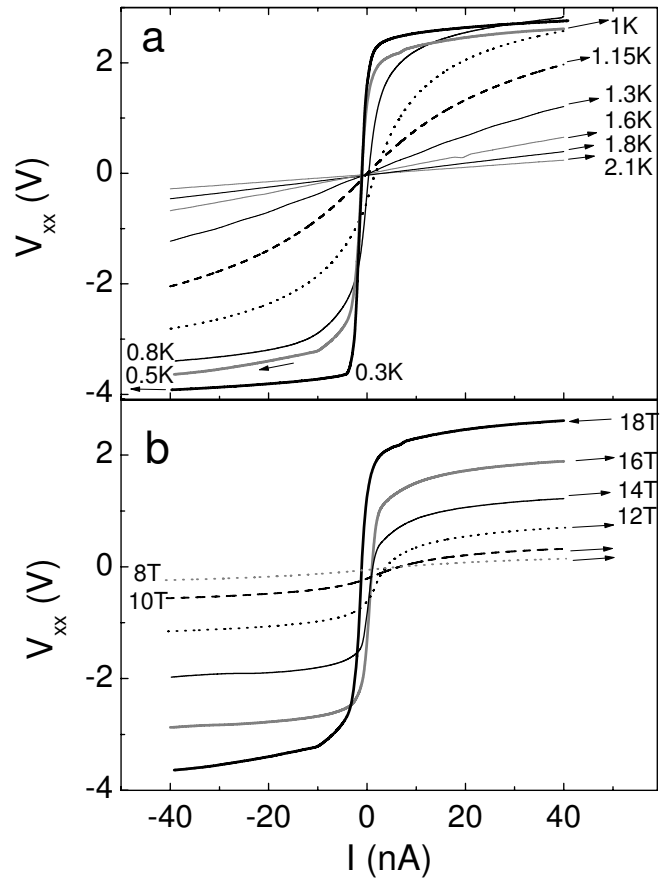}
} \caption{(a) Voltage-current curves for $B=18$ T and different
temperatures. (b) Voltage-current curves for $T=0.55$ K and
different magnetic fields. The ramping speed is 5 nA/min. Arrows
show the ramping direction.
 \label{fig:4}}
\end{figure}

\begin{figure}[hb]
\centerline{
\includegraphics[width=7 cm]{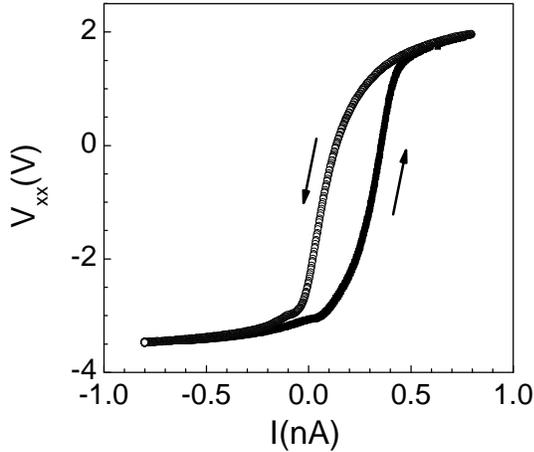}
} \caption{Hysteretic $V-I$ curves measured at $T=0.3$ K and $B=18$
T
  and ramping speed 0.1 nA/min. Arrows show the ramping direction.
 \label{fig:5}}
\end{figure}

\paragraph{Voltage - current curves.}\label{cv}
Shown in Fig.~\ref{fig:4} are $V-I$ curves for different
temperatures and magnetic fields.
One can see that the non-Ohmic behavior starts at very low current,
the $V-I$ curves being asymmetric with respect to the $V$-axis
showing hysteresis at small currents, see Fig.~\ref{fig:5}. The
sample resistance in the hysteretic region depends on the ramping
rate. Note that the voltage-current curves show hysteretic behavior
\textit{only} in the domain of magnetic fields and temperatures
where they are essentially \textit{nonlinear}.

\begin{figure}[ht]
\centerline{
\includegraphics[width=7 cm]{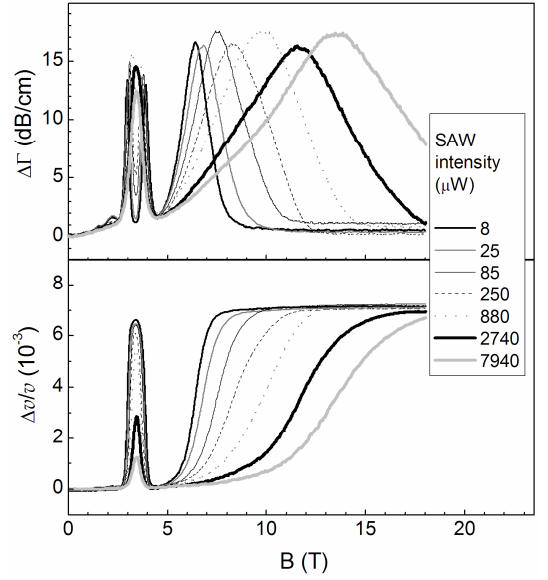}
} \caption{Magnetic field dependence of $\Delta \Gamma$ and $\Delta
v/v (0)$
  at different RF-source powers.  $T=0.3$ K; $f=30$ MHz.
 \label{fig:6}}
\end{figure}

\paragraph{Acoustic properties.} The results of acoustic measurements for
  different SAW intensities are shown in Fig.~\ref{fig:6}.
One can see that in the quantum-limit  region,  $B \ge 4.5$ T,
increase in the SAW intensity acts similarly to an increase in
temperature. Namely, both the attenuation maximum and saturation of
the SAW velocity shift towards large magnetic fields. Similar
behaviors are observed at other SAW frequencies.

\subsection{Data handling} \label{interpretation}

The components $\sigma_{1,2}$ can be found from simultaneous
measurement of $\Delta \Gamma$ and $\Delta v/v(0)$ by solving the
set of equations~\cite{Drichko00}
\begin{eqnarray}
&&   \frac{\Delta \Gamma\,
  (\text{dB/cm})}{8.68kA(k,a,d)}=\frac{\Sigma_1(B)}{[1+\Sigma_2(B)]^2+
  \Sigma_1^2(B)}
\nonumber \\ && \qquad \qquad \qquad \ \qquad
-\frac{\Sigma_1(0)}{[1+\Sigma_1(0)]^2+
  \Sigma_1^2(0)}\, ,  \label{eq:01} \\
&&\frac{v(B)-v(0)}{v(0)A(k,a,d)}=\frac{1+\Sigma_2(B)}{[1+\Sigma_2(B)]^2+
  \Sigma_1^2(B)}
\nonumber \\ && \qquad \qquad \qquad \qquad
-\frac{1+\Sigma_2(0)}{[1+\Sigma_1(0)]^2+
  \Sigma_1^2(0)}  \label{eq:02}
\end{eqnarray}
where
\begin{eqnarray}
&&A(k,a,d)= 110.2b(k,a,d)e^{-2k(a+d)}\, ,  \nonumber \\
&&\Sigma_i=4\pi
  t(a,k,d)\sigma_i/\varepsilon_sv(0)\, ; \nonumber \\
&&
b(k)=(b_1(k)[b_2(k)-b_3(k)])^{-1}
  \, , \nonumber \\
&&t(k,a,d)=[b_2(k)-b_3(k)]/2b_1(k)\, , \nonumber \\
&& b_1(k,a)=(\varepsilon_1+\varepsilon_0)(\varepsilon_s+\varepsilon_0)
\nonumber \\ && \quad \quad \quad
- (\varepsilon_1-\varepsilon_0)
(\varepsilon_s-\varepsilon_0)e^{-2ka}\, ,  \nonumber \\
&&
b_2(k,d)=(\varepsilon_1+\varepsilon_0)(\varepsilon_s+\varepsilon_0)
\nonumber \\ && \quad \quad \quad
+ (\varepsilon_1+\varepsilon_0)
(\varepsilon_s-\varepsilon_0)e^{-2kd}\, ,  \nonumber \\
&&
b_3(k,a,d)=
(\varepsilon_1-\varepsilon_0)(\varepsilon_s-\varepsilon_0)e^{-2ka}
\nonumber \\ && \quad \quad \quad
+(\varepsilon_1-\varepsilon_0)
(\varepsilon_s+\varepsilon_0)e^{-2k(a+d)}\, ,
\end{eqnarray}
$k$ is the SAW wave vector, $d$ is the depth of the 2D-system
layer in the sample, $a$ is the clearance between the sample and the
LiNbO$_3$ surface; $\varepsilon_1$=50, $\varepsilon_0$=1 and
$\varepsilon_s$=11.7 are the dielectric  constants of LiNbO$_3$, of
 vacuum, and of the semiconductor, respectively.

\begin{figure}[ht]
\centerline{
\includegraphics[width=7cm]{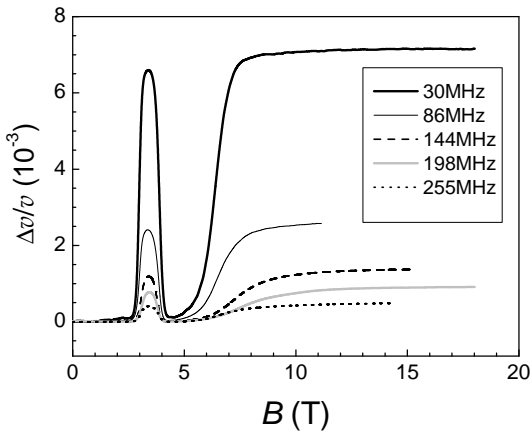}
} \caption{The SAW velocity shift versus magnetic field at different
frequencies at the same sample mounting, $T$=0.3 K.
 \label{fig:VelDiffFreq}}
\end{figure}

Utility of the above
expressions is facilitated by the fact that at $B=0$ the conductance
is metallic, and in the metallic state
$\sigma_1 (\omega)$ is excellently approximated by the static
conductance, $\sigma^{\text{DC}}$, for all relevant frequencies. Thus
we can calibrate the AC response in the absence of magnetic field by
the DC response. For example, for $T=0.3$ K we put
$\sigma_1\vert_{B=0}=\sigma^{\text{DC}} (0) =(1.33\pm 0.02)\times
10^{-4}$ Ohm$^{-1}$. The corresponding value of $\Sigma_1$ for the
relevant frequency range turns out to be much greater than 1. On the
other hand, in the case of metallic conductance one can expect
$\Sigma_2 \ll 1$.~\cite{Drichko00} Thus at $B \to 0$
\begin{equation}
  \label{eq:03}
  \frac{\Delta v (0)}{v(0)A(k,a,d)} \approx \frac{1}{1+\Sigma_1^2} \to
  0\, .
\end{equation}
At $B \to \infty$, both $\Sigma_1$ and $\Sigma_2$ vanish and $\Delta
v/v(0)$ saturates at the value $A(k,a)$. That is exactly what we see
in Fig.~\ref{fig:3}. Furthermore, we can find $A(k,a)$ from the
saturated value of $[\Delta v/v(0)]_{B\to \infty}$. Knowing $k$, we
find from this quantity the thickness of the clearance, $a$. As an
example, at $T=0.3$ K and $f=30$ MHz the saturated value of $\Delta
v/v(0)$ is $7.16\times 10^{-3}$ that corresponds to
$a=4.3\times10^{-5}$ cm. We have checked that this value agrees with
the results for different frequencies - 86, 144, 198, and 255 MHz -
if the sample was not re-installed between the measurements.
Figure~\ref{fig:VelDiffFreq} shows the saturation of the SAW velocity
$\Delta v/v(0)$ in high magnetic fields at different frequencies.
Knowing $a$, $d$, and $\sigma^{\text{DC}}$(0) we can calculate
$\Gamma(0)$, and, respectively, find the absolute value of the SAW
attenuation: $\Gamma(B)= \Delta \Gamma(B) + \Gamma(0)$. This is an
important part of the procedure because in high magnetic fields we
cannot find this quantity directly. Indeed, as one can see from
Fig.~\ref{fig:3}, at large magnetic fields the quantity $\Delta
\Gamma$ is very small and the accuracy of its extraction from the
raw AC data could be insufficient. Having determined the necessary
parameters we then solve the set (\ref{eq:01})-(\ref{eq:02}) for the
quantities $\sigma_1 (\omega)$ and $\sigma_2 (\omega)$.~\cite{end2}

\section{Results} \label{results}
\subsection{QHE regime} \label{QHE regime}

\begin{figure}[ht]
\centerline{
\includegraphics[width=7cm]{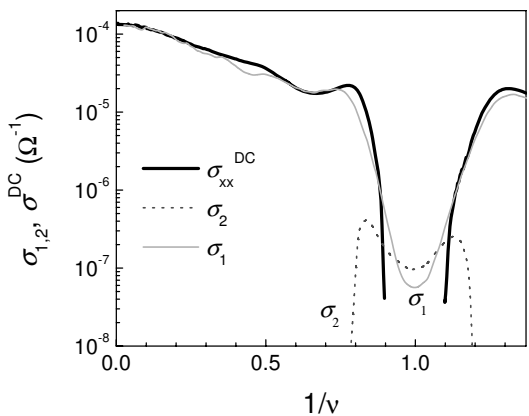}
} \caption{DC conductivity and real \& imaginary components of
AC-conductivity at 30 MHz on the reversed filling factor in the
vicinity of $\nu =1$; $T=0.3$ K.
 \label{fig:Sigmanu1}}
\end{figure}

Figure~\ref{fig:Sigmanu1} illustrates experimental dependences of the
real, $\sigma_{1}$, and imaginary, $\sigma_{2}$, components of the
complex AC conductivity derived from the acoustical measurements using
Eqs.~(\ref{eq:01}) and (\ref{eq:02}), as well as of the DC-conductivity
$\sigma_{xx}^{\text{DC}}=\rho_{xx}/(\rho_{xx}^2+\rho_{xy}^2)$, on the
reciprocal filling factor $1/\nu \propto B$ at $T$=0.3 K in the
vicinity of the filling factor $\nu=1$.
As follows from this figure, outside the vicinity of $\nu =1$ the
values of $\sigma^{\text{DC}}$ and  $\sigma_{1}$ are close. This is
exactly the region of metallic conductance where the hole states are
extended. The situation dramatically changes close to $\nu =1$ where
$\sigma_{1} \gg \sigma^{\text{DC}}$. In addition, in this region the
imaginary component,  $\sigma_{2}$, of the AC
conductivity becomes noticeable.

\begin{figure}[ht]
\centerline{
\includegraphics[width=7cm]{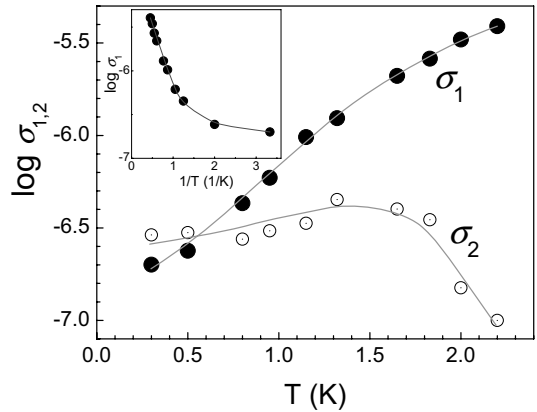}
} \caption{Real and imaginary components of the AC-conductivity
versus
  temperature for $f= 18$ MHz and $\nu$=1. Inset: The Arrhenius plot ($\log
\sigma_1$ versus $1/T$) for the same parameters.
 \label{fig:Sigmanu1T}}
\end{figure}

In Fig.~\ref{fig:Sigmanu1T}, the components $\sigma_{1}$ and
$\sigma_{2}$ are plotted as functions of temperature for $B=3.2$ T
($\nu =1$) and $f=18$ MHz. One can see that the ratio  $\sigma_{1}/
\sigma_{2}$ increases with temperature.  We believe that the
low-temperature region where $\sigma_{2} > \sigma_{1}
> \sigma^{\text{DC}}$ corresponds to the hopping of the holes between
localized states in the  random potential produced by charged
impurities.~\cite{Efros}
We observe that the AC hopping conductance is suppressed as
temperature increases. At highest temperatures ($T=0.8-4.2$ K) the
temperature dependence of  $\sigma_1$ is clearly governed by thermal
activation, $\sigma_1 \propto e^{-\Delta E/T}$, with $\Delta E
\approx g^*\mu_B B$, see inset in Fig.~\ref{fig:Sigmanu1T}.

\subsection{Extreme Quantum Limit} \label{Extreme Quantum Limit}
\begin{figure}[ht]
\centerline{
\includegraphics[width=\columnwidth]{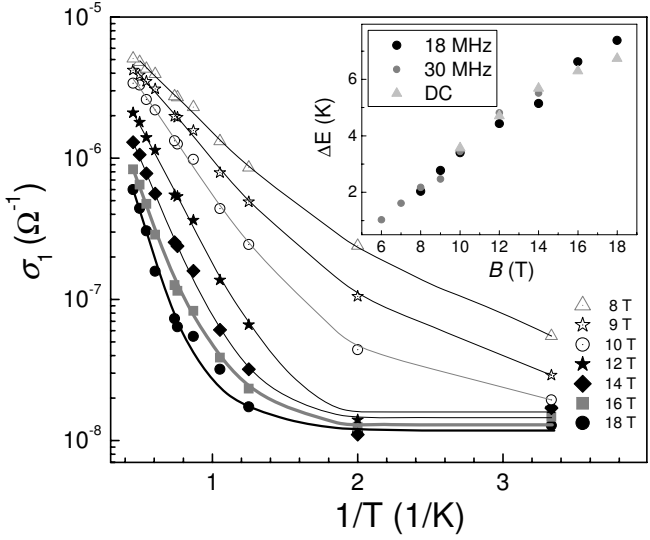}
} \caption{Dissipative AC conductivity, $\sigma_1$,
  versus inverse temperature, $1/T$, for different magnetic
  fields.  $f=18$ MHz, the lines are guides for eye. Inset: magnetic
  field dependence of the activation energy.
 \label{fig:7}}
\end{figure}

\paragraph{Acoustic properties.} Shown in Fig.~\ref{fig:7} is the
dependence of $\sigma_1$ at
frequency $f=18$ MHz on inverse temperature, $1/T$, in different
magnetic fields.
We observe that at $T=0.8-2$ K and $B=8-18$ T the temperature
dependence of $\sigma_1$ is well described by the Arrhenius law,
$\sigma_1 \propto e^{-\Delta E/T}$, with the activation energy,
$\Delta E$, increasing with magnetic field, see inset. In this
domain of temperatures and magnetic fields, for all the measured
frequencies (18-87 MHz) $\sigma_1 \approx \sigma^{\text{DC}}$
(measured for the current $I=0.5$ nA).

Both quantities depend on temperature according to the Arrhenius
law, the activation energies being very close, see Fig.~\ref{fig:7},
inset.
 We
believe that the observed activation dependences of $\sigma_1$ and
$\sigma^{\text{DC}}$ are due to magnetic freeze-out of the holes
from the extended states of the $0$th Landau level to the states
localized nearby the Fermi level.
At low temperatures and large magnetic fields ($T < 0.8$ K, $B \ge
11$ T) the temperature dependence of $\sigma_1$ becomes very weak.

\begin{figure}[ht]
\centerline{
\includegraphics[width=8.7 cm]{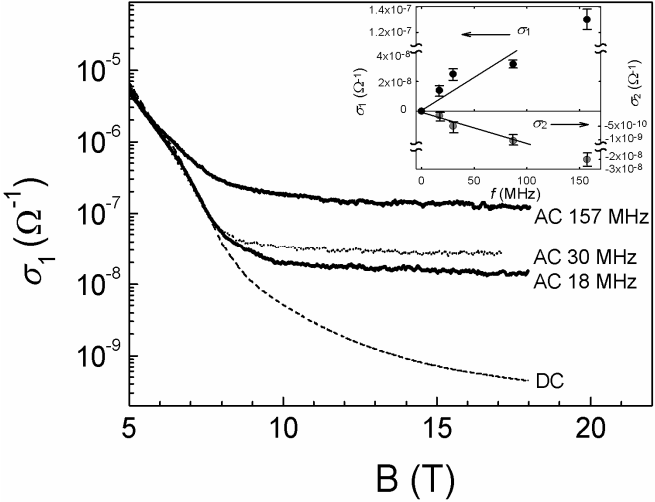}
} \caption{Real part of the AC conductivity, $\sigma_1$ at different
  frequencies and DC conductivity measured at 0.5 nA  versus magnetic
  field; $T=0.3$ K.   Inset: frequency
  dependences of $\sigma_1$ and $\sigma_2$ at $B=18$ T, $T=0.3$ K.
 \label{fig:8}}
\end{figure}

Magnetic field dependences of $\sigma_1$ at $T=0.3$ K and
different frequencies as well as the DC conductivity measured with
the current excitation of 0.5 nA are shown in Fig.~\ref{fig:8}.

At $B> 8$ T,  $\sigma^{\text{DC}} /\sigma_1 \ll 1$; this ratio can
be less than 0.1. In addition, both the temperature and magnetic
field dependences of $\sigma^{\text{DC}}$ cannot be accounted for by
conventional expressions for hopping conductance. The magnetic field
dependence of $\sigma_1$ weakens as magnetic field increases; at
$B\ge 12$ T the dependence is almost absent.
Inset of Fig.~\ref{fig:8} illustrates the
frequency dependences of $\sigma_{1,2}$ at $B=18$ T and  $T=0.3$ K.
The low-frequency part of these dependences are \textit{linear}:
 $\sigma_{1,2} (\omega)
\propto \omega$. However, the slopes have different signs -- the
imaginary part of the conductivity, $\sigma_2$, is
\textit{negative}.
The experimental points for $f=157$ MHz cannot be allowed for by the
linear dependence, note that there is a break in the scale.

\begin{figure}[ht]
\centerline{
\includegraphics[width=\columnwidth]{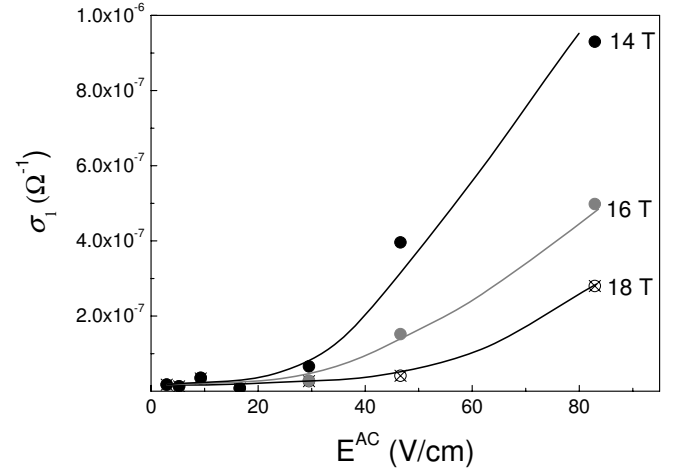}
} \caption{Dependence of $\sigma_1$ on the SAW amplitude,
  $E^{\text{AC}}$, at different magnetic fields. $T=0.5$ K, $f=30$
  MHz. The lines are guides for eye.
 \label{fig:9}}
\end{figure}

Shown in Fig.~\ref{fig:9} are dependences of $\sigma_1$ on the
amplitude,  $E^{\text{AC}}$, of the electric filed produced by the SAW
in the 2DHG.
This amplitude is calculated from the expression~\cite{DrichkoNonlin}
\begin{equation}
  \label{eq:05}
  \left(E^{\text{AC}}\right)^2=\frac{32\pi K^2}{v(0)}\,
  \frac{W }{l} e^{-2k(a+d)}
  \frac{kb(k,a,d)(\varepsilon_1+\varepsilon_0)}{(1+\Sigma_2)^2+\Sigma_1^2} \, .
\end{equation}
Here $W$ is the acoustic power carried by the SAW, $l$ is the
SAW aperture.
The rest of notations is the same as for Eqs.~(\ref{eq:01}) and
(\ref{eq:02}). Though the accuracy of this expression is not high one
can still conclude that in strong magnetic field the nonlinear
behavior starts at higher SAW intensities.

\paragraph{DC measurements}
Contrary to the case of AC conductance, the non-Ohmic behavior of
the static $V-I$ curves, Fig.~\ref{fig:4}, starts as a threshold.
The threshold electric field, $E_t$, increases with the magnetic
field. At $B=18$ T and $T=0.3$ K, $E_t\approx 9$ V/cm. The threshold
positions depend on the ramping speed of the current. However, even
at the lowest ramping speed (0.02~nA/min) and smallest currents ($I
< 0.1$ nA) the $V-I$ curves remain non-Ohmic. Thus one can conclude
that the mechanisms of nonlinear behavior at zero and finite
frequency are significantly different.

\begin{figure}[ht]
\centerline{
\includegraphics[width=\columnwidth]{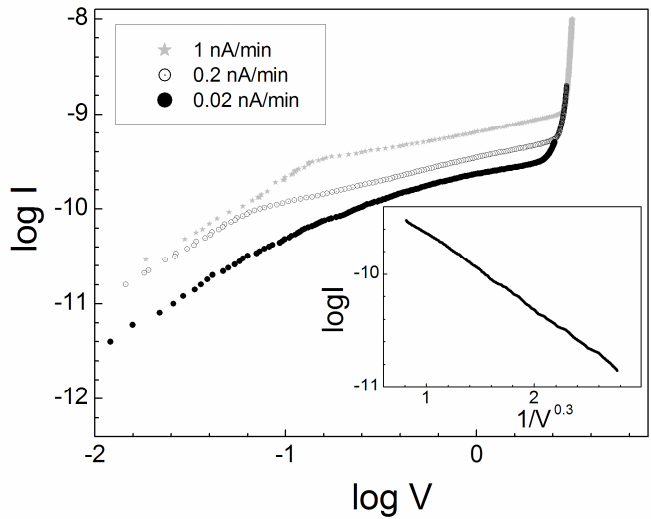}
} \caption{$\log$-$\log$ plot of the $I-V$-curves with different
current sweep rate at $B=18$, $T=0.3$ K. Inset: Logarithm of the
current versus $V^{-0.3}$ for the curve taken on the ramping speed
0.02 nA/min
 \label{fig:11}}
\end{figure}

As we already mentioned, the $V-I$ curves depend on the current
ramping rate. To analyze the static nonlinear $I-V$ curves it is
convenient to plot them in the $\log$-$\log$ scale. Shown in
Fig.~\ref{fig:11} is the $\log$-$\log$ plot of $V-I$ curves at
$B=18$, $T=0.3$ K and different current sweep rate.
The curve with the lowest ramping speed, 0.02 nA/min, can be
fitted as
\begin{equation}
  \label{eq:06}
  I \propto e^{-C/V^\alpha}
\end{equation}
where $C$ is a constant while $\alpha=0.3-0.4$, see inset.

\section{Discussion and conclusion} \label{discussion}

{}From the experimental results and their analysis we conclude that
the conduction mechanisms in $p$-SiGe heterostructures are different
in different domains of temperature and magnetic field.

In the absence of magnetic field,  $B=0$, the conductance shows
metallic behavior, see Fig.~\ref{fig:3}. In relatively weak magnetic
fields, close to the filling factor $\nu =1$, there is a
temperature-driven crossover from the hopping conductance at low
temperatures to the thermal activation at higher temperatures.
Indeed, at low temperatures ($T < 0.6$ K) both $\sigma_1 (\omega)$
and $\sigma_2 (\omega)$ relatively weakly depend on temperature, and
the inequality $\sigma_2 > \sigma_1 > \sigma^{DC}$ holds. This
behavior is compatible with the predictions for hopping between the
localized states created by the random impurity potential. The
results in this domain of the magnetic field and temperature can be
allowed for using conventional models for hopping conductance in 2D
systems.~\cite{Efros} At $T \gtrsim 1$ K, $\sigma_2$ gets much less
than $\sigma_1$ and finally becomes not measurable,  the temperature
dependence of $\sigma_1$ crosses over to the thermal activation
(with magnetic field dependent activation energy). This behavior is
typical for the conductance driven by extended states.

The extreme quantum limit occurs in very strong magnetic fields
where $\nu< 1$. As follows from Fig.~\ref{fig:2}, the DC resistance
at $B\gtrsim 4.5$ T ($\nu \lesssim 0.8$) becomes rapidly increasing
with magnetic field.  As the temperature increases, the temperature
dependence of $\sigma_1$ crosses over to thermal activation with
magnetic field dependent activation energy, see Fig.~\ref{fig:7}.
This behavior can be understood as a metal-to-insulator transition
(MIT) driven by the magnetic field. A similar transition was
observed by many authors in $p$-Si/SiGe heterostructures, see,
e.\,g., Ref.~\onlinecite{Coleridge}. A crossover from metallic to
insulating behavior manifests itself also in the acoustic
properties, as seen in Fig.~\ref{fig:3} -- at $B \gtrsim 4.5$ T both
$\Gamma$ and $\Delta v/v(0)$ cease to depend on temperature. With
increase of temperature $\sigma_1$ crosses over to the Arrhenius
law, see Fig.~\ref{fig:7}. This law holds at $T\gtrsim 1$ K for
$B=18$ T and $T\gtrsim 0.5$ K for $B=10$ T.

At the lowest temperature (0.3 K) and $4.5 \, \text{T} < B< 7\,
\text{T}$, $\sigma_1
>\sigma_2>0$, $\sigma_1$ being  almost independent of
frequency, see Fig.~\ref{fig:8}.
We believe that in this interval of magnetic fields the process of
magnetic freeze-out takes place: occupation of extended states
exponentially decreases with temperature, the activation energy
increasing with magnetic field.

At very high magnetic field and low temperatures, $B
> 14$ T and   $ T < 0.8$ K, we face rather unusual behavior.
One could expect that at low temperatures and in very strong
magnetic field the carriers are localized by a random potential, therefore
both DC and AC conductance are due to single-particle hopping
between the localized states. This would lead to an exponentially
small Ohmic DC conductance, $\sigma^{\text{DC}}$, which should be
much less than $\sigma_1 (\omega)$. In addition, the predicted
$\sigma_2$ is positive, and it should significantly exceed
$\sigma_1$ in the whole studied frequency range.\cite{Drichko00}
Such a behavior was indeed observed in the vicinity of $\nu =1$,
i.\,e., in the integer quantum Hall effect regime. However, the
experimental results for low temperatures and very high magnetic
field strongly differ from these predictions. Indeed, the $I-V$
curves are nonlinear even for very small voltages. At larger
voltages there is a pronounced increase in the current when the
voltage exceeds some threshold value. The $I-V$ curves can be
described by Eq.~(\ref{eq:06}), which is typical for the creep
motion of pinned charge density wave, or Wigner
glass.\cite{Ioffe87,Feigelman89,Blatter94} In the same region of
temperatures and magnetic fields, the AC response is linear up to
the electric field amplitudes, $\sim 20$ V/cm; $\sigma_1 \propto
\omega$ is almost independent of temperature and magnetic field. The
nonlinear effects are qualitatively similar to those caused by
heating of carriers by the electric field created by the SAW. In
addition, the imaginary part of the linear-response conductivity is
\textit{negative}, and
 $ |\sigma_2| \ll \sigma_1$ that contradicts
 to the predictions of the model allowing for the single-particle
 hopping.

We believe that the observed experimental results can be explained
assuming that at low temperatures and in strong magnetic field the
holes form a \textit{Wigner glass}, i.\,e., a Wigner crystal distorted
by disorder-induced pinning. Indeed, at small voltage and low
temperature the pinned Wigner crystal should  behave as an insulator.
At finite temperature, parts of the Wigner glass experience
correlated hops between different pinned states leading to the
charge transfer. This process is similar to the \textit{creep} of
dislocations~\cite{Ioffe87} or pinned vortices in type-II
superconductors.~\cite{Feigelman89,Blatter94}  The law (\ref{eq:06})
and hysteretic $I-V$ curves are typical for the creep motion.

The dynamic response of weakly pinned Wigner crystal at  not too
small frequencies is dominated by the collective
excitations~\cite{Fogler99,Fogler00} where an inhomogeneously
broadened absorption line (the so-called pinning mode)
appears.~\cite{Fukuyama77} It corresponds to collective vibrations
of correlated segments of the Wigner crystal around their
equilibrium positions formed by the random pinning potential. The
mode is centered at some disorder- and magnetic-field-dependent
frequency, $\omega_p$; its width being determined by a complicated
interplay between different collective excitations in the Wigner
crystal. There are modes of two types: transverse (magnetophonons)
and longitudinal (magnetoplasmons). The latters include fluctuations
in electron density. An important point is that pinning modifies
both modes, and the final result depends on the strength and
correlation length, $\xi$, of the random potential. Depending in the
strength and  correlation length of the random potential, the
frequency, $\omega_p$ may either increase, or decrease with magnetic
field. As follows from experiments in GaAs, $\omega_p \approx
10^{10}$ s$^{-1}$ ($f \approx 1.5$ GHz). Consequently, in our
acoustic experiments $\omega/\omega_p \ll 1$. Hence, we were not
able to observe the peak in the frequency dependence of the
attenuation and can discuss only its low-frequency tail. The ratio
$\omega_p/\omega_c$ can be arbitrary. Depending on the interplay
between the ratio  $\omega_p/\omega_c$ and the ratio $\eta \equiv
\sqrt{\lambda/\beta}$ between the shear ($\beta$) and bulk
($\lambda$) elastic moduli one can specify two regimes where the
behaviors of $\sigma^{\text{AC}}$ are different:
\begin{equation} \label{regimes}
(a) \ 1 \ll \omega_c/\omega_{p0} \ll
\eta, \quad (b) \ 1 \ll \eta
  \ll \omega_c/\omega_{p0}     \, .
\end{equation}
Here $\omega_{p0}$ is the pinning frequency at $B=0$. As a result, the
variety of different behaviors is very rich.
 Assuming $\xi \gg
l_B=(\hbar c/eB)^{1/2}$ one can cast the expression for
$\sigma_{xx}(\omega)$ from Ref.~\onlinecite{Fogler00} into the form
\begin{equation}
  \sigma_{xx}(\omega)=i\frac{e^2n\omega}{m^*\omega_{p0}^2}\frac{1+iu(\omega)}
{[1+iu(\omega)]^2   -(\omega
  \omega_c/\omega_{p0}^2)^2}\, , \label{sigma1}
\end{equation}
where the function $u(\omega)$ is different for regimes (a) and (b).
Note that the above equation differs from that of
Ref.~\onlinecite{Fogler00} by replacement $\omega \to -\omega, \
u\to -u$ since the sign of $\omega$ used in the Fourier transform in
Ref.~\onlinecite{Fogler00} is opposite to that used in
Eq.~(\ref{eq:02}).

Below we will focus on the regime (b) since only this regime seems to
be compatible with experimental results. Then
\begin{equation}
  \label{eq:fb}
  u(\omega) \sim \left\{ \begin{array}{lll}
(\omega/\Omega)^{2s}, & \omega \ll \Omega\, , &\quad  (b1)\\
\text{const}, & \Omega \ll \omega \ll \omega_c \, .& \quad (b2) \end{array}
\right.
\end{equation}
Here $\Omega \sim \omega_{p0}^2\eta/\omega_c$, while $s$ is some
critical exponent. According to Ref.~\onlinecite{Fogler00}, $s=3/2$.
For the regime $(b2)$ of Eq.~(\ref{eq:fb}) we have
\begin{equation}
  \frac{\sigma_{xx}(\omega)}{\sigma_0}= \frac{\omega}{\omega_{p0}}\cdot
\frac{i(1+iu)}{(1+iu)^2-(\eta
  \omega /\Omega)^2}  \label{sigma1b}
\end{equation}
where $\sigma_0 \equiv e^2p/m^*\omega_{p0}$. The result can be cast
in the form
\begin{eqnarray}
  \sigma_1&=&\sigma_0 u \frac{\omega}{\omega_{p0}} \frac{1+u^2+(\eta
  \omega/\Omega)^2}{[1+u^2+(\eta \omega/\Omega)^2]^2-(2\eta
  \omega/\Omega)^2}, \label{resigma}\\
 \sigma_2&=&-\sigma_0 \frac{\omega}{\omega_{p0}} \frac{1+u^2-(\eta
  \omega/\Omega)^2}{[1+u^2+(\eta \omega/\Omega)^2]^2-(2\eta
  \omega/\Omega)^2} . \label{imsigma}
\end{eqnarray}
The above prediction is qualitatively compatible with the experimental
results if one assumes
\begin{equation}
  \label{eq:001}
  u \gg \omega \omega_c/\omega_{p0}^2 \gg 1\, .
\end{equation}
Then
\begin{equation}\label{lsigma}
 \sigma_1 \approx \sigma_0\frac{\omega}{ u } \, , \quad
 \sigma_2 =-\sigma_0\frac{\omega}{u^2}
\end{equation}
and $\sigma_1/|\sigma_2|=u \gg 1$. As follows from the experimental
data (see inset of Fig.~\ref{fig:8}), the components $\sigma_2 < 0$,
both $\sigma_1$ and $\sigma_2$ almost proportional to frequency, and
their ratio is $\sigma_1/|\sigma_2| \approx 40$, i.\,e., greater
than 1. In addition, the experimentally measured $\sigma_1$ and
$\sigma_2$ are almost independent of magnetic field and temperature
in the domain where we expect formation of the Wigner crystal.

The regime (\ref{lsigma}) requires the inequalities (\ref{eq:001}) and
(\ref{eq:fb}, $b$2) to be met simultaneously.
 These conditions impose restriction on the
frequency $\omega_{p0}$ of the collective mode. Indeed, they are
both  valid at the if $\omega_{p0} \approx (1.2-1.5) \times 10^{10}$
s$^{-1}$. In this way one can determine the pinning frequency in the
absence of  magnetic field. At $\sigma_1/|\sigma_2| \approx 40$ and
$B>14$ T, the inequality (\ref{eq:001}) is valid only for
frequencies 18-90 MHz. At frequency $f=157$ MHz, the ratio  $\omega
\omega_c/\omega_{p0}^2$ is much greater than 40. That can explain
why the linear frequency dependence of $\sigma_1$ and $\sigma_2$
does not hold up to the highest experimental frequency. Taking into
account low accuracy in determination of $\sigma_2$, the agreement
between theory and experiment can be considered as satisfactory.

Assuming that $ \omega_{p0} =1.5 \times 10^{10}$ s$^{-1}$, $u=40 $ and
using Eq.~(\ref{sigma1}) one can estimate the
concentration of the holes participating in formation of the Wigner
crystal in the extreme quantum limit. The estimate yields a value,
which is approximately by 2 orders of magnitude  smaller, than the
hole density in the absence of magnetic field ($8.2 \times 10^{10}$
cm$^{-2}$). Apparently, in the very strong magnetic field
most of carriers is captured by impurities and only small part is
involved in formation of the Wigner glass. Unfortunately, we cannot
compare this result with any data on
Hall resistance, $\rho_{xy}$, since we were not able to
measure this quantity in the extreme quantum limit.

Knowing the frequency $\omega_{p0}$, we can estimate the typical
correlation length, $L$, of a pinned Wigner crystal.
Following Ref.~\onlinecite{corr-length} we have
\begin{equation}
  \label{freqpin}
  \omega_{p0}=c_t (2\pi/L)\, ,
\end{equation}
where $c_t=\sqrt{\beta/p m^*}$ is the velocity of transverse phonons
in the Wigner crystal; $m^*$ is the hole effective mass;
$\beta=0.245 e^2 p^{3/2}/\varepsilon_s$ is the shear elastic
modulus.

Assuming the effective hole density at $T=0.3$ K and $B=18$ T to be
$p \approx 10^9$ cm$^{-2}$ (i.\, e., two orders lower then 2DHG
sheet density)  one estimates the correlation length as $L \approx 4
\times 10^{-4}$ cm. The  lattice constant of the Wigner crystal $a_W
\equiv (\pi p)^{-1/2} \approx 4 \times 10^{-5}$ cm. Furthermore,
with such $p$ the ratio of the hole-hole interaction energy to the
Fermi energy $\kappa=E_{hh}/E_F \approx 70$. Thus, the inequalities
which are necessary conditions for the Wigner crystal formation --
$L \gg a_W$ and $\kappa \gg$1 -- are met.

Though the above estimates produce reasonable numbers one should not
overestimate their accuracy. The point is that, along with the
``pinning'' modes, there exist localized ``soft'' ones. They appear in
the places where pinning is weak.\cite {Fogler99} One can expect
that these modes can also contribute to the AC conductance similarly
as it happens in structural glasses.~\cite{Hunklinger} Moreover, the
frequency dependence of their contribution to $\sigma_1(\omega)$  is
close to linear, while the contribution to $\sigma_2$ is small.
Unfortunately, the density of these soft modes is unknown, therefore
we are not able to estimate these contributions quantitatively.

To conclude the discussion, we believe that  $p$-Si/SiGe
heterostructures demonstrate crossover between different mechanisms of
DC and AC conductance - from metallic conductance at $B=0$ through
hopping in the integer quantum Hall effect regime to the pinned Wigner
crystal (Wigner glass)
in the extreme quantum limit (at $B>14$ T and $T<0.8$ K). The
conclusion regarding formation of the Wigner glass is supported by the
behavior of complex AC conductance showing small negative imaginary
part compatible with the predictions of Ref.~\onlinecite{Fogler00}
for the  pinning mode of a Wigner crystal, as well as by a
creep-like nonlinear behavior and hysteretis of DC conductance. This
conclusion agrees with that of Ref.~\onlinecite {Pudalov94}
based on DC measurements.

\acknowledgments

We are grateful to V.M. Pudalov and V. I. Kozub for useful discussions;
to E. Palm and T. Murphy for their help with the experiments performed at the
NHMFL.

The work is supported by grants of the Presidium of the Russian
Academy of Science, the Program of Branch of Physical Sciences
"Spintronika", St. Petersburg Scientific Center of RAS 2007; NSF
Cooperative Agreement No. DMR-0084173, State of Florida; NHMFL
In-House Research Program. The work of YG, and VV was supported by
the U. S. Department of Energy Office of Science through contract
No. DE-AC02-06CH11357 and by Norwegian Research Council through
USA-Norway bilateral agreement.


\begin{thebibliography}{99}
\bibitem{qhe}  \textit{"The Quantum Hall effect"}, ed. by R. E. Prange
  and S. M. Girvin (Springer-Verlag, 1987).
\bibitem{LL} L. D. Landau, E. M. Lifshits, \textit{Kvantovaya Mechanika},
v.3, 1963, p.704, Moskva, Fizmatgiz.

\bibitem{insul} O. Viehweger and K. B. Efetov, J. Phys. Condens. Matter, 2,
7049 (1990); S. Kivelson, D.-H. Lee, and S.-C. Zhang, Phys. Rev. B
46, 2223 (1992).

\bibitem{Efros} A. L.~Efros, Zh. Eksp. Teor. Fiz. {\bf 89}, 1834
(1985) [JETP {\bf 89}, 1057 (1985)].

\bibitem{Drichko00} I. L. Drichko, A. M. Diakonov, I. Yu. Smirnov,
  Y. M. Galperin, and A. I. Toropov,  \prb \textbf{62}, 7470 (2000).

\bibitem{Wigner34} E. Wigner,
Phys. Rev. \textbf{46}, 1002 (1934).

\bibitem{Grimes79} C. C. Grimes and G. Adams, \prl \textbf{42},795
  (1979).

\bibitem{Pudalov94} V. M. Pudalov, in: \textit{Phys. of Quantum
    Sol. of Electrons}, 124 (1994), Int. Press.

\bibitem{Lozovik75} Y. E. Lozovik and  V. I. Yudson, JETP
  Lett. \textbf{22}, 11 (1975) [Pis'ma Zh. Eksp. Teh. Fiz \textbf{22}, 26 (1975)].

\bibitem{Jiang92} H. W. Jiang, H. L. Stormer, D .C. Tsui, L. N.
  Pfeiffer, and K. W. West, \prb \textbf{44}, 8107 (1991).

\bibitem{Dolgopolov92} V. T. Dolgopolov, G. V. Kravchenko,
  A. A. Shashkin, S. V. Kravchenko, Phys. \prb \textbf{46}, 13303
  (1992).

\bibitem{Podor01} B. P\"od\"or, Gy.  Kov\'acs, G. Rem\'enyi, I. G.
  Savel'ev, and
  S. V.~Novikov,  Inorg. Mat. \textbf{37}, 439 (2001).

\bibitem{Fogler99} M. M. Fogler and D. A. Huse, \prb \textbf{59}, 2120 (1999).

\bibitem{Ioffe87} L .B. Ioffe and V. M. Vinokur,J. Phys. C
  \textbf{20}, 6149 (1987).

\bibitem{Feigelman89} M. V. Feigel'man, V.B. Geshkenbein,
  A. I. Larkin and V. M. Vinokur, \prl \textbf{63}, 2303 (1989).

\bibitem{Blatter94}G. Blatter, M. V. Feigel'man, V. B. Geshkenbein,
  A. I. Larkin, and V. M. Vinokur, \rmp \textbf{66}, 1125 (1994).

\bibitem{Williams91} F. I. B. Williams, P. A. Wright, R. G. Clark,
E. Y. Andrei, G. Deville, D. C. Glattli, O. Probst, B.  Etienne, C.
Dorin, C.T . Foxon, and J. J. Harris, \prl \textbf{66}, 3285 (1991).

\bibitem{Ye02} P. D. Ye, L. W. Engel, D. C. Tsui, R .M. Lewis,
  L. N. Pfeiffer, and K. West, \prl \textbf{89}, 176802 (2002).

\bibitem{Li00} C. C. Li, J. Yoon, L.W. Engel, D. Shakhar, D. C. Tsui
  and M. Shayegan, \prb \textbf{61}, 10905 (2000). C. C. Li, L.W. Engel, D. Shakhar, D. C. Tsui
  and M. Shayegan, \prl \textbf{79}, 1353 (1997).

\bibitem{Yong06} P. Yong, G. Chen, G. Sambamdamurthy, Z. H. Wang,
  R. M.  Lewis, L .W. Engel, D. C. Tsui, P. D. Ye, L. N. Pfeiffer, and
  K. West, Nature Physics \textbf{2}, 452 (2006).

\bibitem{Paalanen92} M. A. Paalanen, R. L. Willett, P.B. Littlewood,
  R.R. Ruel, K. West, L. N. Pfeiffer, and D. J. Bishop, \prb
  \textbf{45},11342 (1992).


\bibitem{Fedina} L. Fedina, O. I. Lebedev, G. Van Tendeloo,
J. Van Landuy, O. A. Mironov and E. H. C. Parker, \prb \textbf{61},
10336 (1999).

\bibitem{Agan} S. Agan, O. A. Mironov, M. Tsaousidou, T. E. Whall,
  E. H. C. Parker, and P. N. Butcher, Microelectronic Engineering
\textbf{51-52}, 527 (2000).

\bibitem{Coleridge} P. T. Coleridge,
Sol. St. Com. \textbf{127}, 777 (2003).

\bibitem{end1} Small ($\le 10$ \%) difference in the peak positions
  can be attributed to the fact that the samples for DC and acoustic
  were cut from the
neighbouring area of the heterostructure.
\bibitem{end2} Note that the accuracy in determining of the
clearance $a$ strongly influences the obtained values of $\sigma_2$
while $\sigma_1$ is less sensitive to $a$.

\bibitem{DrichkoNonlin} I. L. Drichko, A. M. Diakonov, I. Yu. Smirnov,
A. I. Toropov, Semiconductors \textbf{34}, 422 (2000) [FTP
\textbf{34}, 436 (2000)].



\bibitem{Fogler00} M. M. Fogler and D. A. Huse, \prb \textbf{62}, 7553
  (2000).

\bibitem{Fukuyama77} H. Fukuyama and P. A. Lee, \prb \textbf{17}, 535
  (1977); \prb \textbf{18}, 6245 (1978).

\bibitem{corr-length} B. G. A. Normand, P. B. Littlewood and A. J.
Millis, \prb \textbf{46}, 3920 (1992).


\bibitem{Hunklinger} S. Hunklinger and W. Arnold, in \textit{Physical
    Acoustics}, edited by W. P. Mason and R. N. Thurston (Academic, New
  York, 1976), Vol. XII, p. 155;
 S. Hunklinger and A. K. Raychaudhuri, in \textit{Progress in Low
    Temperature Physics}, edited by D.F. Brewer (Elsevier, Amsterdam,
    1986), Vol. IX, p. 267.


\end{thebibliography}
\end{document}